\documentclass{JINST}

\usepackage{graphicx}

\usepackage{amsmath}
\usepackage{pifont}
\newcommand{\m}{\operatorname{m}}             
\newcommand{\cm}{\operatorname{cm}}           
\newcommand{\mm}{\operatorname{mm}}           
\newcommand{\psymu}{\ensuremath{\text{\small\Pisymbol{psy}{109}}}}  
\newcommand{\mum}{\operatorname{\psymu m}}    
\newcommand{\s}{\operatorname{s}}             
\newcommand{\mus}{\operatorname{\psymu s}}    
\newcommand{\ns}{\operatorname{ns}}           
\newcommand{\ps}{\operatorname{ps}}           

\title{Performance of the Fast Beam Conditions Monitor BCM1F of CMS in the first running periods of LHC}

\author{R. S. Schmidt$^{a,b,}$\thanks{Corresponding author.}~, A. J. Bell$^{c,g}$, E. Castro$^b$,
  R. Hall-Wilton$^{c,d}$, M. Hempel$^a$, W. Lange$^b$, W. Lohmann$^{a,b}$, S. M\"uller$^{c,e}$,
  V. Ryjov$^c$, D. Stickland$^{c,f}$, and R. Walsh$^b$\\
\llap{$^a$}Brandenburgische Technische Universit\"at, 03046 Cottbus, Germany\\
\llap{$^b$}DESY, 15738 Zeuthen \& 22607 Hamburg, Germany\\
\llap{$^c$}CERN, 1211 Gen\`eve 23, Switzerland\\
\llap{$^d$}University of Wisconsin, Madison, WI 53706, USA\\
\llap{$^e$}Karlsruher Institut f\"ur Technologie, 76049 Karlsruhe, Germany\\
\llap{$^f$}Princeton University, Princeton, NJ 08544, USA\\
\llap{$^g$}Universit\'e de Gen\`eve, 1211 Gen\`eve, Switzerland\\
  E-mail: \email{ringo.schmidt@desy.de}}

\abstract{The Beam Conditions and Radiation Monitoring System, BRM, is
  implemented in CMS to protect the detector and provide an interface
  to the LHC. Seven sub-systems monitor beam conditions and the
  radiation level inside the detector on different time scales. They
  detect adverse beam conditions, facilitate beam tuning close to CMS,
  and measure the doses accumulated in different detector components.
  Data are taken and analysed independently of the CMS data
  acquisition, displayed in the control room, and provide inputs to
  the trigger system and the LHC operators. In case of beam conditions
  dangerous to the CMS detector, a beam abort is induced.

  The Fast Beam Conditions Monitor, BCM1F, is a flux counter close to
  the beam pipe inside the tracker volume. It uses single-crystal CVD
  diamond sensors, radiation-hard FE electronics, and optical signal
  transmission to measure the beam halo as well as collision products
  bunch by bunch. The system has been operational during the
  initiatory runs of LHC in September 2008. It works reliably since
  the restart in 2009 and is invaluable to CMS for everyday LHC
  operation. A characterisation of the system on the basis of data
  collected during LHC operation is presented.}

\keywords{Control and monitor systems online; Data acquisition concepts; Radiation-hard electronics}

\begin{document}

\section{The CMS Beam Conditions and Radiation Monitoring System, BRM}

The Compact Muon Solenoid (CMS)~\cite{CMS} is a multi-purpose detector
experiment at interaction point (IP)~5 of the Large Hadron Collider
(LHC) at CERN.  Beam losses in the LHC may cause serious harm to
detector components, so their advent must be detected in order to
avert damages of the detector. A monitoring system is needed that
allows diagnosis of adverse beam conditions and can initiate beam
aborts or shut down vulnerable detectors, if necessary.

\begin{figure}[b]
  \centering
  \includegraphics[width=0.85\textwidth]{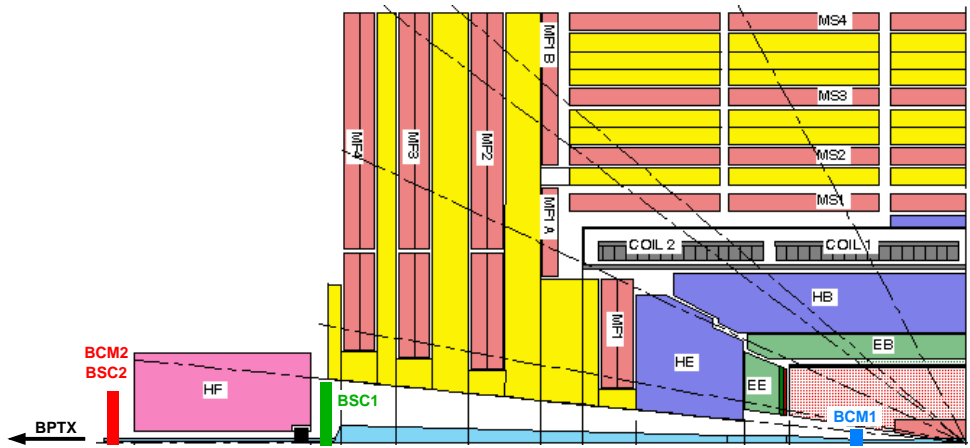}
  \caption{Quadrant of a longitudinal section of CMS with the IP in the bottom right corner. The surrounding layers are the tracker (red), the electromagnetic (green) and the hadronic (blue) calorimeter, both consisting of barrel and endcap, the coil (grey), and the return yoke with interspersed muon chambers (yellow/red). HF is the hadronic forward calorimeter. Along the beam pipe, the positions of BRM sub-systems are indicated (starting at IP, cf.\ Table~\protect\ref{tab_BRM}): BCM1 at the blue bar, BSC1 at the green bar, BCM2/BSC2 at the red bar, and BPTX (out of scope). RADMONs and passive elements are distributed throughout the detector.}
  \label{fig_CMS}
\end{figure}

Throughout the CMS detector, the beam conditions and the radiation
level are monitored by seven sub-systems working on different time
scales~\cite{BRM}.  Locations of these systems are indicated in
Figure~\ref{fig_CMS}, and some of their specifications are listed in
Table~\ref{tab_BRM}~\cite{Sensors}.  These systems are operated
independently of the LHC power supply and the CMS data acquisition and
must be active whenever there might be beam in the LHC.

\begin{table}[htb]
  \centering
  \caption{Some specifications of the BRM sub-systems, which are ordered by increasing time resolution. RADMON stands for Radiation Monitor, BCM for Beam Conditions Monitor (F:~fast, L:~leakage), BSC for Beam Scintillation Counter, and BPTX for Beam Pick-up Timing Experiments. The IP is located at $z=0$.}
  \begin{tabular}{ ccccc }
    Sub-system  & Sensor        & Location              & Time resolution       & Function   \\
    \hline\hline
    Passives    & TLD           & in CMS and cavern     & long-term             & monitoring \\
    RADMON      & RadFET + SRAM & around CMS            & $1\s$                 & monitoring \\
    BCM2        & pCVD diamond  & $z = \pm 14.4\m$      & $40\mus$              & protection \\
    BCM1L       & pCVD diamond  & $z = \pm 1.8\m$       & $\approx 5\mus$       & protection \\
    BSC         & scintillator  & $z=\pm (10.9,14.4)\m$ & $\ns$                 & monitoring \\
    BCM1F       & sCVD diamond  & $z = \pm 1.8\m$       & $\ns$                 & mon./prot. \\
    BPTX        & beam pickup   & $z = \pm 175\m$       & $200\ps$              & monitoring \\
  \end{tabular}
  \label{tab_BRM}
\end{table}

\section{The Fast Beam Conditions Monitor, BCM1F}

\paragraph{Conceptual design}
Four modules, consisting of sensor, pre-amplifier, and optical driver,
are arranged around the beam pipe on either detector side at distances
of $4.5\cm$ from the beam axis and~$\pm 1.8\m$ from the IP.  They
measure the flux of beam halo particles as well as of collision products,
thereby providing \emph{CMS Background\,1} for the LHC control.
Particularly radiation-hard components are needed in this situation.
Additional requirements are low power dissipation and excellent time
resolution in order to detect single relativistic charged particles
with a time resolution better than the time between bunch crossings~\cite{BCM1F-NSS}.

\begin{figure}[b]
  \centering
  \includegraphics[width=\textwidth]{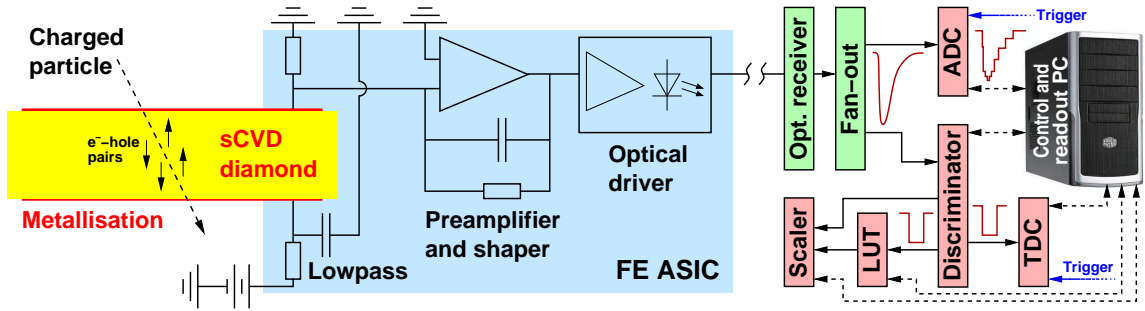}
  \caption{Schematic of the BCM1F readout chain (not to scale). (Left) Front-end module, including sensor, preamplifier, and optical driver; through the lowpass filter, test pulses can be injected. (Right) Signal processing units of the back-end in the counting room.}
  \label{fig_ASIC}
\end{figure}

\paragraph{Readout chain and data acquisition}
The sensors are single-crystal chemical vapour deposition (sCVD)
diamonds of the size ${5\mm} \times {5\mm} \times 400\mum$. They are
metallised on both sides and operated as solid state ionisation
chambers, as illustrated in Figure~\ref{fig_ASIC}.  Charge-sensitive,
radiation-hard amplifier ASICs of the type JK16~\cite{JK16} collect
the charges induced and shape proportional signals. These are
transmitted to the counting room as optical signals, converted back to
electrical signals, and processed and analysed by several back-end
components.  The analogue signals are fed into a discriminator that
supplies scalers and time-to-digital converters (TDC) with logical
signals. An analogue-to-digital converter (ADC) samples the input
voltages to obtain digitised signal images. 


Data acquired by all these readout devices are processed immediately.
Relevant results are displayed in the control room and passed on to
the central CMS data acquisition.  Raw data are stored permanently on
disk for offline analyses.

\begin{figure}[htb]
  \centering
  \includegraphics[clip,trim=20 15 30 20,width=0.48\linewidth]{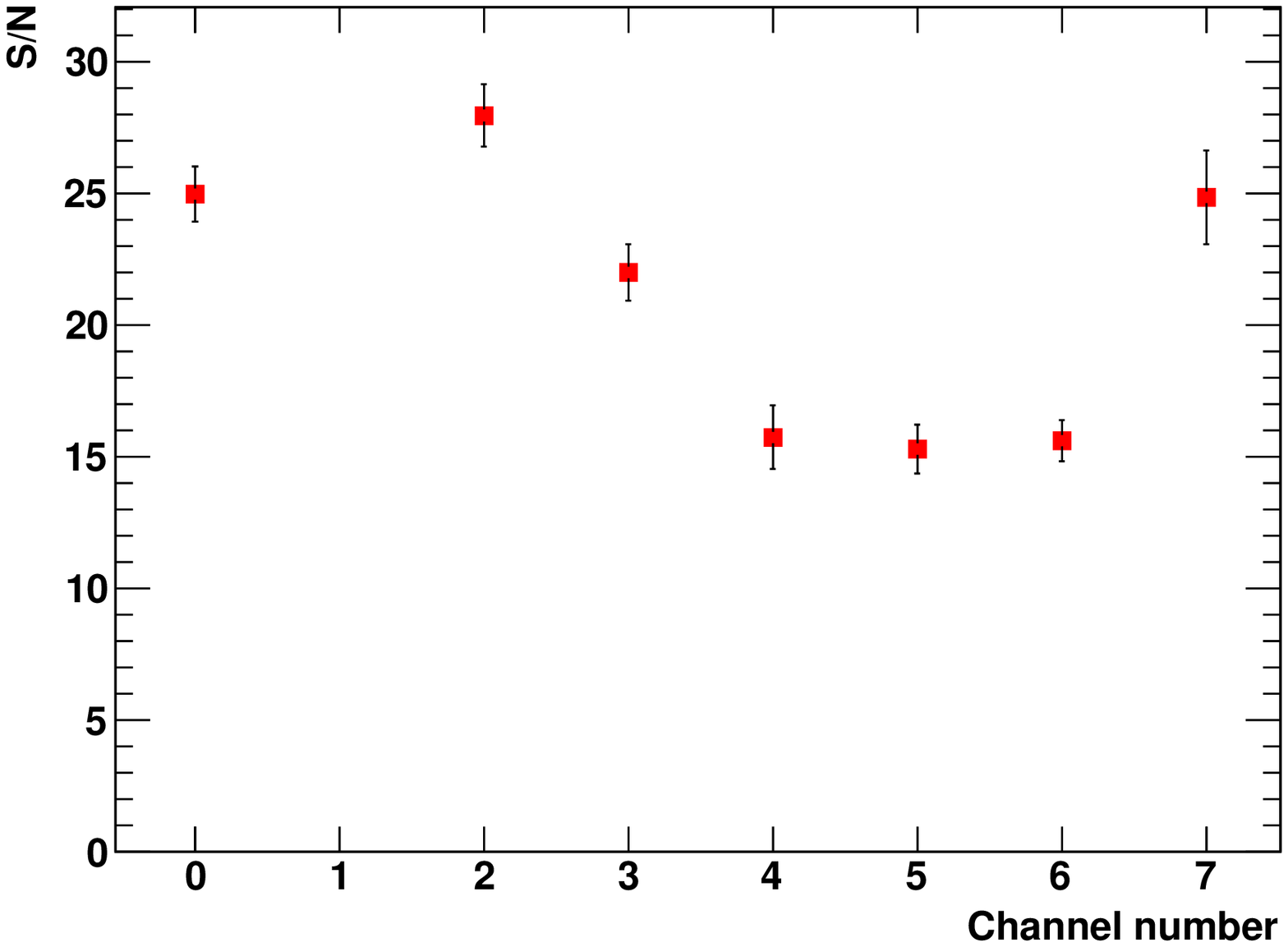}
  \hfill
  \includegraphics[clip,trim=20 15 30 20,width=0.48\linewidth]{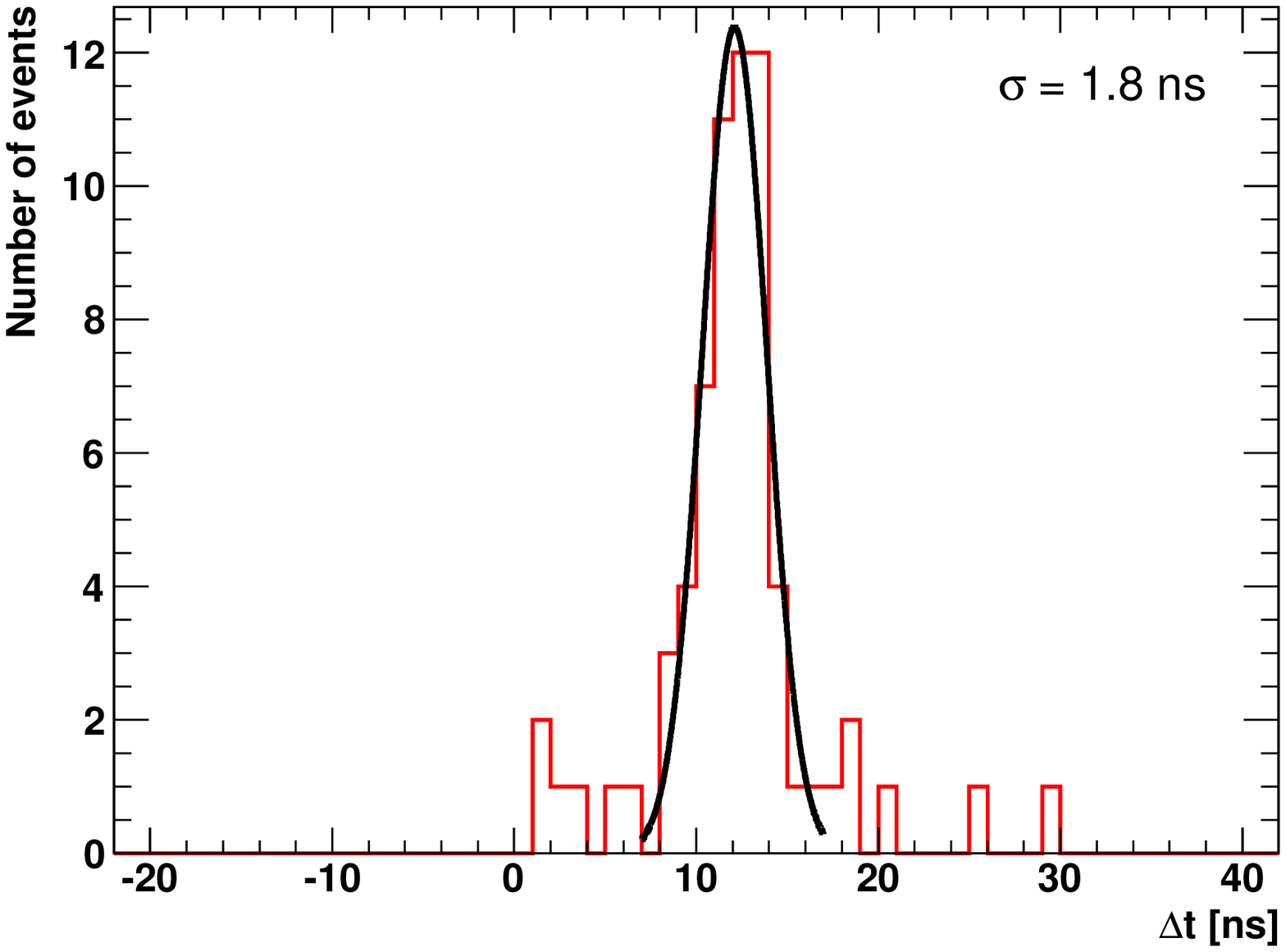}
  \caption{Performance of BCM1F with first LHC beams in September 2008. (Left) Signal-to-noise ratio of the channels. (Right) Time resolution of the system.}
  \label{fig_perform}
\end{figure}

\paragraph{Results obtained with first beams}
Digitised signals from the ADC are used to understand the performance
of the system. It entirely meets the expectations as could be proved
already in the first runs of LHC in 2008. The plots in
Figure~\ref{fig_perform} show the signal-to-noise ratio of the
detectors\footnote{Channel 1 had a faulty cable in 2008. Meanwhile,
  this has been replaced, and Channel 1 delivers data as well.} and a
time resolution measurement~\cite{BCM1F-NIM}.  The latter was taken
while only one beam was circulating and shows the time
difference between signals from sensors at opposite detector planes
and equal azimuthal positions.  The mean value of $12\ns$ corresponds
perfectly to the time of flight of relativistic particles for $3.6\m$,
and a standard deviation of $\sigma = 1.8\ns$ results in a time
resolution of $1.3\ns$.

\section{Performance and Prospects of BCM1F}

\subsection{Signal Spectra from the ADC}

\begin{figure}[htb]
  \parbox[c]{0.49\textwidth}{\centering%
    \includegraphics[clip,trim=0 0 45 25,width=\linewidth]{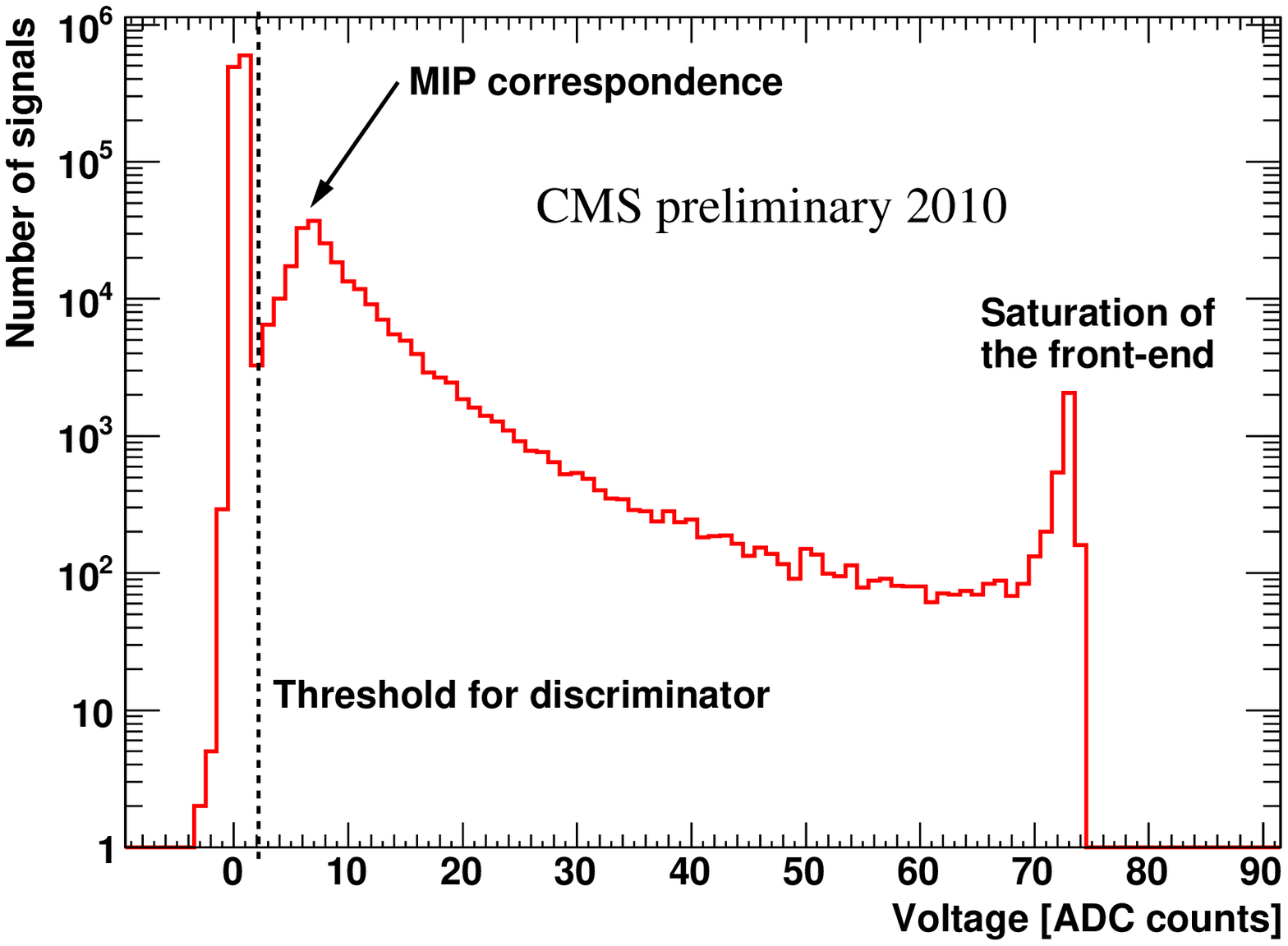}}%
  \hfill\parbox[c]{0.46\textwidth}{\centering%
    \includegraphics[clip,angle=270,width=\linewidth]{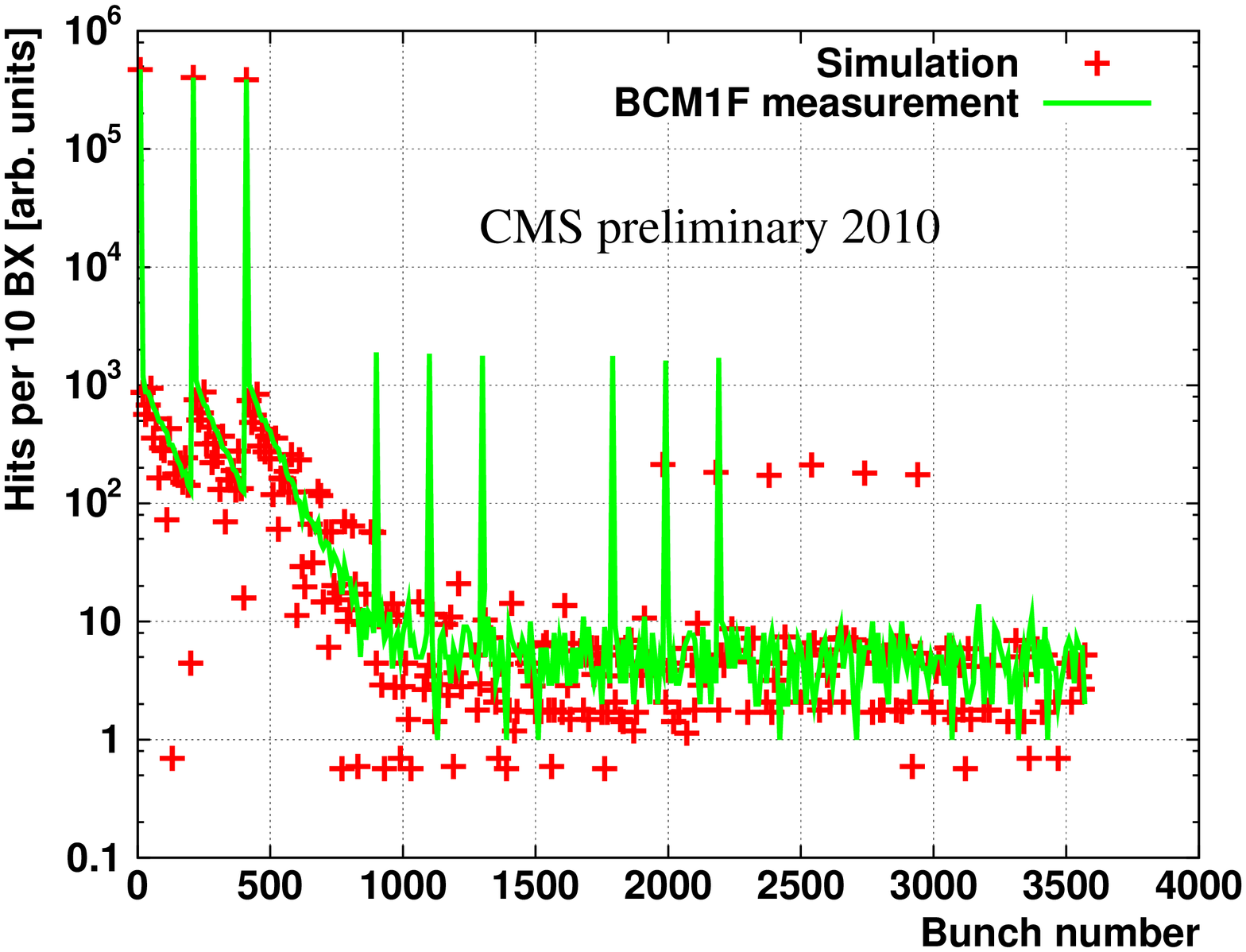}}\\%
  \parbox[t]{0.49\textwidth}{\centering%
    \caption{Pulse height spectrum obtained from the ADC to find discriminator thresholds from the local minimum between the pedestal (left peak) and the MIP signals. The third peak at high voltages is caused by signals in saturation.}
    \label{fig_ampl}}%
  \hfill\parbox[t]{0.46\textwidth}{\centering%
    \caption{Bunch structure over the LHC orbit monitored with the TDC (green line) and simulated using FLUKA (red markers). A statistical effect not related to bunches lead to six outlying markers between bunch numbers 2000 and 3000.}
    \label{fig_bunches}}%
\end{figure}

ADC data have been invaluable for the commissioning of the system.
They facilitate maintenance and can deliver online information as well.
The pulse height spectrum of signals taken in a run with colliding
bunches is displayed in Figure~\ref{fig_ampl} for ADC input channel 0.
The trigger used for this measurement is an analogue sum of all BCM1F
channels.  Therefore, a fraction of the events contains the baseline
in this channel, which accumulates to the pedestal peak at about zero
pulse height. The maximum position of the signal peak is considered
the pulse height for single relativistic particles (MIP
correspondence)\footnote{The energy of relativistic particles will
  usually be larger than the energy of minimum ionising particles
  (MIPs), thus their energy loss will be slightly larger than that of
  MIPs. This effect is ignored here.}.  The local minimum between
those peaks is used to determine the optimal threshold to be set in
the discriminator.

The third peak as well as the adjacent cut at the upper end of the
spectrum result from a limitation of the front-end electronics.  At a
pulse height of about ten times the MIP amplitude, the output signal
of the laser driver goes into saturation.

\subsection{Timing Information from the TDC}

The arrival times of hits in BCM1F are recorded with the TDC. 
Given the spacing of $24.95 \ns$ between potential bunch crossings, 
time can be converted into a bunch number within the LHC orbit.
An example of the count rate as a function of the
bunch number is displayed in Figure~\ref{fig_bunches}. 
The first three peaks, corresponding to colliding bunches,
exhibit very high count rates and long tails up
to the $\mus$ range.  Both effects are caused by collision products.
The peaks around bunch number 1000 and 2000, respectively, represent
the three non-colliding bunches of each beam.
No tails are observed here.

A simulation has been performed~\cite{Bunches} using
FLUKA~\cite{FLUKA} to understand the reasons for the long tails of
colliding bunches.\footnote{The simulation is normalised to the data
  and does not include non-colliding bunches.} The results agree well
with the data, as can be seen in Figure~\ref{fig_bunches}, and confirm
that, in addition to collision products and beam halo, also delayed
signals from electrons, photons, and neutrons contribute to the count
rates.


\subsection{Online Analysis with Scalers}

\begin{figure}[htb]
  \centering
  \includegraphics[clip,width=0.75\textwidth]{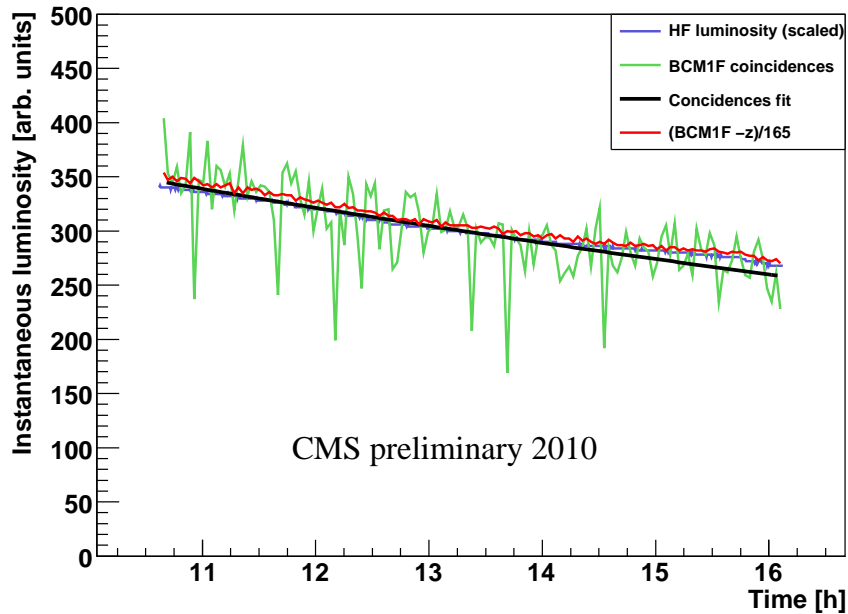}
  \caption{Comparison of the luminosity measured by HF with count rates from BCM1F.}
  \label{fig_lumi}
\end{figure}

The output signals of the discriminator are also fed into scalers and a
freely programmable logic unit, see Figure~\ref{fig_ASIC}.  The unit's
FPGA is programmed as a look-up table (LUT) to match pairs of BCM1F
modules opposing with respect to the IP.  Coincidences of signals in
such back-to-back modules are assumed to be caused by elastic
scattering in the collisions. Corresponding logical coincidence
signals are counted by the scalers as well.  In Figure~\ref{fig_lumi},
the overall coincidence rate and the luminosity measured by the
hadronic forward calorimeter (HF) are plotted against time.  The HF
result (blue line) has been scaled to the BCM1F coincidence rate
(green line), which has been fit (black line) in order to compare the
slopes.  In addition, the scaled count rate of BCM1F on the $-z$ side
is shown (red line).  We observe a reasonable agreement between the
three lines, demonstrating the potential of BCM1F to be used as a fast
luminosity monitor.

\begin{thebibliography}{9}

\bibitem{CMS} The CMS Collaboration, 
  \emph{The CMS experiment at the CERN LHC}, 
  \jinst{3}{2008}{S08004}.

\bibitem{BRM} Alan J. Bell on behalf of the BRM group,
  \emph{Beam \& Radiation Monitoring for CMS},
  {2008 \emph{IEEE Nuclear Science Symposium Conference Record}, pp.\ 2322--2325}.

\bibitem{Sensors} A. Macpherson,
  \emph{Beam Conditions and Radiation Monitoring at the {LHC} Experiments},
  {2006 \emph{Proceedings LHC Project Workshop, Chamonix XV}, p. 198}.

\bibitem{BCM1F-NSS} R. Hall-Wilton, W. Lange, A. Macpherson, V. Ryjov, and R.L. Stone,
  \emph{Fast beam conditions monitor (BCM1F) for CMS},
  {2008 \emph{IEEE Nuclear Science Symposium Conference Record}, pp.\ 3298--3301}.

\bibitem{JK16} J. Kaplon and W. Dabrowski,
  \emph{Fast CMOS binary front end for silicon strip detectors at LHC experiments},
  {2005 \emph{IEEE Transactions on Nuclear Science 52(6)},  pp.\ 2713--2720}.

\bibitem{BCM1F-NIM} A.J. Bell, E. Castro, R. Hall-Wilton, W. Lange, W. Lohmann, A. Macpherson, M. Ohlerich, N. Rodriguez, V. Ryjov, R.S. Schmidt, and R.L. Stone,
  \emph{Fast beam conditions monitor BCM1F for the CMS experiment},
  {2010 \emph{Nuclear Instruments and Methods in Physics Research A614}, pp.\ 433--438}.
  arXiv:0911.2480

\bibitem{Bunches} Steffen M\"uller on behalf of the CMS collaboration,
  \emph{Impact of beam-induced backgrounds for the CMS Pixel and other inner radii detectors (Simulation and Data)},
  {poster at the International Workshop on Semiconductor Pixel Detectors for Particles and Imaging, September, 6--10, 2010, proceedings to be published in \emph{Nuclear Instruments and Methods in Physics Research A, Special issue: PIXEL 2010}}.\\
  Steffen M\"uller,
  \emph{Design, Commissioning and Performance of the CMS Beam Condition Monitor~2 and Simulation Studies of the Radiation Environment near CMS at LHC},
  {2010 Ph.\,D.~Thesis in Preparation, CERN\,/\,KIT Karlsruhe}.

\bibitem{FLUKA} A. Fasso, A. Ferrari, J. Ranft, and P.R. Sala,
  \emph{FLUKA: a multi-particle transport code},
  {2005 CERN-2005-10, INFN/TC\_05/11, SLAC-R-773}.

\end{thebibliography}
\end{document}